\documentclass[12pt]{article}
\usepackage{graphicx}
\usepackage[cp1251]{inputenc}

\begin{document}
 \noindent {\footnotesize\it
   Astronomy Letters, 2020, Vol. 46, No 2, pp. 131--143.}
 \newcommand{\dif}{\textrm{d}}

 \noindent
 \begin{tabular}{llllllllllllllllllllllllllllllllllllllllllllll}
 & & & & & & & & & & & & & & & & & & & & & & & & & & & & & & & & & & & & & \\\hline\hline
 \end{tabular}

  \vskip 0.5cm
  \centerline{\bf\large Kinematics of T Tauri Stars Close to the Sun from }
  \centerline{\bf\large the Gaia DR2 Catalogue}
   \bigskip
  \bigskip
  \centerline
 {V.V. Bobylev\footnote [1]{e-mail: vbobylev@gaoran.ru}}
  \bigskip

  \centerline{\small\it Pulkovo Astronomical Observatory, Russian Academy of Sciences,}

  \centerline{\small\it Pulkovskoe sh. 65, St. Petersburg, 196140 Russia}
 \bigskip
 \bigskip
 \bigskip

 {
{\bf Abstract}---The spatial and kinematic properties of a large sample of young T Tauri stars from the solar neighborhood 500 pc in radius have been studied. The following parameters of the position ellipsoid have been determined from the most probable members of the Gould Belt: its sizes are $350\times270\times87$~pc and it is oriented at an angle of $14\pm1^\circ$ to the Galactic plane with a longitude of the ascending node of $297\pm1^\circ$. An analysis of the motions of stars from this sample has shown that the residual velocity ellipsoid with principal semiaxes $\sigma_{1,2,3}=(8.87,5.58,3.03)\pm(0.10,0.20,0.04)$~km s$^{-1}$ is oriented at an angle of $22\pm1^\circ$ to the Galactic plane with a longitude of the ascending node of $298\pm2^\circ$. It has been established that much of the expansion effect (kinematic $K$ effect) typical for Gould Belt stars, 5--6 km s$^{-1}$ kpc$^{-1}$, can be explained by the influence of a Galactic spiral density wave with a radial perturbation amplitude $f_R\sim5$ km s$^{-1}$.
  }


 \subsection*{INTRODUCTION}
The Gould Belt is a fairly flat system with semiaxes
of $\sim350\times250\times50$~pc, with the direction of its
semimajor axis being near $l=40^\circ$ (Efremov 1989;
P\"oppel 1997, 2001; Torra et al. 2000; Olano 2001).
The plane of its symmetry is inclined to the Galactic
plane approximately at $i=18^\circ$. The longitude of the
ascending node is $l_\Omega=280^\circ$. The Sun is at a distance
of $\sim40$ pc from the line of nodes. The system's
center lies at a heliocentric distance of 100--150 pc
in the second Galactic quadrant. The estimate of the
direction to the center $l_0$ depends on the sample age
and, according to various published sources, ranges
from $130^\circ$ to $180^\circ$. The spatial distribution of stars
is highly nonuniform---a noticeable drop in density is
observed within $\approx80$~pc of the center, i.e., the entire
system has the shape of a doughnut. The well-known
open star cluster $\alpha$~Per with an age of $\sim35$ Myr lies
near the center of this doughnut. A number of nearby
OB associations (de Zeeuw et al. 1999) and open star
clusters (Piskunov et al. 2006; Bobylev 2006), dust
(Dame et al. 2001; Gontcharov 2019) and molecular
(Perrot and Grenier 2003; Bobylev 2016) clouds
belong to the Gould Belt; a giant neutral hydrogen
cloud called the Lindblad ring (Lindblad 1967, 2000)
is associated with it.

We know about the kinematic properties of the Gould Belt from an analysis of the motions of young massive O- and B-type stars (Torra et al. 2000), young open star clusters (Piskunov et al. 2006; Bobylev 2006; Vasilkova 2014), and molecular clouds (Perrot and Grenier 2003; Bobylev 2016). In particular, evidence of expansion and intrinsic rotation of this system has been found. Using the Scorpius–Centaurus association closest to the Sun (on a scale of $\sim150$~pc) as an example, Sartori et al. (2003) showed the absence of differences in distribution and kinematics between massive and low-mass (T Tauri) stars of comparable age. On a larger scale ($\sim1$~kpc in diameter) a kinematic analysis of T Tauri stars has not yet been performed due to the absence of necessary measurements. With the appearance of the Gaia DR2 catalogue (Brown et al. 2018; Lindegren et al. 2018), it has become possible to select tens of thousands of such stars (Zari et al. 2018) that belong to known associations closely related to the Gould Belt. These include the Scorpius–Centaurus, Orion, Vela, Taurus, Cepheus, Cassiopeia, and Lacerta associations.

An expansion of individual OB associations (Blaauw 1964), groupings of young associations close to the Sun (Torres et al. 2008), samples of young massive OB stars (Torra et al. 2000), and a large complex of young open star clusters (Piskunov et al. 2006; Bobylev 2006) has been noticed in the Gould Belt region. There is no certainty in the question of what center or line the expansion
originates from, because the effect manifests itself as a dependence of the velocities $U$ and $V$ on coordinates $x$ and $y.$ Bobylev (2014) suggested that much of the Gould Belt expansion could be explained by the influence of a spiral density wave. A practical allowance for the effect, apparently, has not yet been
made and, therefore, the results of this approach are of great interest.

The goal of this paper is to determine the spatial and kinematic properties of a large sample of T Tauri stars from the Gaia DR2 catalogue selected by Zari
et al. (2018). Such an analysis suggests a study of the system's spatial orientation, a confirmation of the system's expansion and intrinsic rotation typical for the Gould Belt, and an analysis of the residual stellar
velocities.

 \section*{DATA}
In this paper we use the compilation by Zari et al. (2018) that contains more than 40 000 T Tauri stars selected from the Gaia DR2 catalogue by
kinematic and photometric data. These stars are within 500 pc of the Sun, because the restriction on the sample radius $\pi>2$ milliarcseconds (mas) was
used. They were selected by proper motions through
an analysis of the smoothed distribution of points
on the $\mu_\alpha\cos\delta\times\mu_\delta$  plane using the restriction on
the tangential stellar velocity
 $\sqrt{\mu^2_\alpha\cos\delta+\mu^2_\delta}<40$~km s$^{-1}$.

The following three subsamples of T Tauri stars are
presented in the catalogue by Zari et al. (2018):

(i) PMS1 that includes 43 719 stars within the outermost contour constructed when smoothing the points on the $\mu_\alpha\cos\delta\times\mu_\delta,$ plane and, therefore, this sample contains the largest number (compared to the
two remaining ones) of background objects;

(ii) PMS2 that contains 33 985 stars within the
second contour on the $\mu_\alpha\cos\delta\times\mu_\delta,$ plane;

(iii) PMS3 that contains 23 686 stars within the
third contour and, therefore, they are the most probable
members of the kinematic grouping (Gould Belt).

In addition, there is a sample of early-type stars
from the Gaia DR2 catalogue located in the upper
main sequence on the Hertzsprung–Russell (H--R)
diagram designated as UMS. It contains 86 102 stars
with an absolute magnitude $M_{G,0}$ less than 3.5$^m.$ In
the opinion of Zari et al. (2018), this sample includes
stars of spectral types O, B, and A.

The line-of-sight velocities in the catalogue by Zari et al. (2018) were taken from various sources, in particular, from the Gaia DR2 catalogue. However,
the stars with line-of-sight velocities are much fewer
than the stars with proper motions. Figure 1 presents
the H–R diagram constructed from the stars of the
UMS and PMS3 samples. On this diagram the stars
with measured line-of-sight velocities are marked;
the stars with line-of-sight velocity measurement errors
no more than 5 km s$^{-1}$ were taken. It can be
seen that there are few very young stars and stars with
measured line-of-sight velocities in the UMS sample,
while in the PMS3 sample, on the contrary, the stars
with measured line-of-sight velocities are relatively
more luminous and evolutionally most advanced, because
they are close to the main sequence on the H--R diagram.

As shown by Zari et al. (2018), the stars of all the samples presented by them, PMS1, PMS2, PMS3, and, to a lesser degree, UMS, have a close spatial
association with the Gould Belt.

 \begin{figure} {\begin{center}
 \includegraphics[width=140mm]{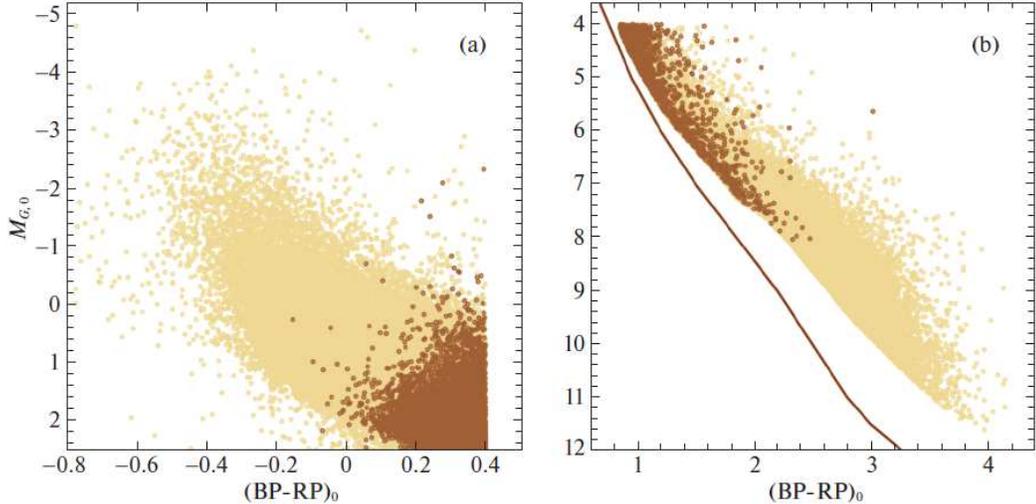}
 \caption{
(Color online) The H–R diagram constructed from the UMS (a) and PMS3 (b) stars, the dark circles mark the stars with measured line-of-sight velocities, the solid line marks the main sequence.}
 \label{f1}
 \end{center} } \end{figure}

 \section*{METHODS}
We use a rectangular coordinate system centered on the Sun in which the $x$ axis is directed toward the Galactic center, the $y$ axis is in the direction of
Galactic rotation, and the $z$ axis is directed toward the north Galactic pole. Then, $x=r\cos l\cos b,$ $y=r\sin l\cos b$ and $z=r\sin b.$

We know three stellar velocity components from observations: the line-of-sight velocity $V_r$ and two tangential velocity components,$V_l=4.74r\mu_l\cos b$ and $V_b=4.74r\mu_b,$ directed along the Galactic longitude $l$ and latitude $b,$ respectively, expressed in km s$^{-1}$. Here, the coefficient 4.74 is the ratio of the number of kilometers in an astronomical unit to the number of seconds in a tropical year and $r=1/\pi$ is the stellar heliocentric distance in kpc that we
calculate via the stellar parallax $\pi$ in mas. The proper motion components $\mu_l\cos b$ and $\mu_b$ are expressed in mas yr$^{-1}$.

For each star the velocities $U, V,$ and $W$ can be calculated via the components $V_r, V_l,$ and $V_b,$ where $U$ is directed from the Sun toward the Galactic center, $V$ is in the direction of Galactic rotation, and $W$ is
directed to the north Galactic pole:
 \begin{equation}
 \begin{array}{lll}
 U=V_r\cos l\cos b-V_l\sin l-V_b\cos l\sin b,\\
 V=V_r\sin l\cos b+V_l\cos l-V_b\sin l\sin b,\\
 W=V_r\sin b                +V_b\cos b.
 \label{UVW}
 \end{array}
 \end{equation}
These velocities can be determined only for those stars for which both line-of-sight velocities and proper motions have been measured.

Let us estimate what stellar line-of-sight velocity errors must be in our sample for them to be comparable to the tangential velocity errors. In the
Gaia DR2 catalogue the mean parallax errors for bright stars $(G<15^m)$ lie within the range 0.02--0.04 mas, while for faint stars $(G =20^m)$ they reach
0.7 mas. Similarly, the proper motion errors range
from 0.05 mas yr$^{-1}$ for bright stars $(G<15^m)$ to
1.2 mas yr$^{-1}$ for faint ones $(G=20^m).$ If we take
a proper motion error of 0.1 mas yr$^{-1}$, then the
tangential velocity error at a sample boundary of
0.5 kpc will be $4.74\times0.5\times0.1=0.2$ km s$^{-1}$, while
for the extreme case, for a proper motion error of 1 mas yr$^{-1}$, the tangential velocity error at the sample boundary will be $4.74\times0.5\times1=2.4$ km s$^{-1}$. Thus, it is desirable to use the stellar line-of-sight velocities with their random measurement errors less than
2.4 km s$^{-1}$.

 \subsection*{Residual Velocity Formation}
When forming the residual velocities, we take into account primarily the peculiar solar velocity, $U_\odot,$ $V_\odot$ and $W_\odot$. Since the diameter of the solar neighborhood considered by us is 1 kpc, the influence of the differential Galactic rotation should also be taken into account. Finally, it is interesting to take into account the influence of the Galactic spiral density wave. The
expressions for a full allowance for the listed effects are
 \begin{equation}
 \begin{array}{lll}
 V_r=V^*_r-[-U_\odot\cos b\cos l-V_\odot\cos b\sin l-W_\odot\sin b\\
 +R_0(R-R_0)\sin l\cos b\Omega^\prime_0
 +0.5R_0(R-R_0)^2\sin l\cos b\Omega^{\prime\prime}_0\\
 + \tilde{v}_\theta \sin(l+\theta)\cos b
 - \tilde{v}_R \cos(l+\theta)\cos b],
 \label{EQU-1}
 \end{array}
 \end{equation}
 \begin{equation}
 \begin{array}{lll}
 V_l=V^*_l-[U_\odot\sin l-V_\odot\cos l-r\Omega_0\cos b\\
 +(R-R_0)(R_0\cos l-r\cos b)\Omega^\prime_0
 +0.5(R-R_0)^2(R_0\cos l-r\cos b)\Omega^{\prime\prime}_0\\
 + \tilde{v}_\theta \cos(l+\theta)
 + \tilde{v}_R\sin(l+\theta) ],
 \label{EQU-2}
 \end{array}
 \end{equation}
  \begin{equation}
 \begin{array}{lll}
 V_b=V^*_b-[U_\odot\cos l\sin b + V_\odot\sin l \sin b-W_\odot\cos b\\
 -R_0(R-R_0)\sin l\sin b\Omega^\prime_0
 -0.5R_0(R-R_0)^2\sin l\sin b\Omega^{\prime\prime}_0\\
 - \tilde{v}_\theta \sin(l+\theta)\sin b
 + \tilde{v}_R \cos(l+\theta)\sin b ],
 \label{EQU-3}
 \end{array}
 \end{equation}
where $V^*_r,V^*_l,V^*_b$ on the right-hand sides of
the equations are the original, uncorrected velocities,
while $V_r, V_l,$ and $V_b$ on the left-hand sides are the
corrected velocities with which we can calculate the
residual velocities $U, V,$ and $W$ based on relations (1),
$R$ is the distance from the star to the Galactic rotation
axis, $R^2=r^2\cos^2 b-2R_0 r\cos b\cos l+R^2_0.$ The
distance $R_0$ is assumed to be $8.0\pm0.15$ kpc. We
take the specific values of the peculiar solar velocity,
$(U_\odot,V_\odot,W_\odot)=(11.1,12.2,7.3)$ km s$^{-1}$, according
to the definition by Sch\"onrich et al. (2010). We use
the following kinematic parameters:
 $\Omega_0=28.71\pm0.22$ km s$^{-1}$ kpc$^{-1}$,
 $\Omega^{\prime}_0=-4.100\pm0.058$ km s$^{-1}$ kpc$^{-2}$, and
 $\Omega^{\prime\prime}_0=0.736\pm0.033$ km s$^{-1}$ kpc$^{-3}$, where $\Omega_0$
is the angular velocity of Galactic rotation at the
distance $R_0,$ the parameters
 $\Omega^{\prime}_0$ and $\Omega^{\prime\prime}_0$ are the
corresponding derivatives of this angular velocity. These parameters were determined by analyzing a sample of young open star clusters with the proper
motions, parallaxes, and line-of-sight velocities calculated from Gaia DR2 data (see Bobylev and Bajkova 2019a).

We can find two velocities, $V_R$ directed radially
away from the Galactic center and the velocity $V_\theta$
orthogonal to it in the direction of Galactic rotation,
based on the following relations:
 \begin{equation}
 \begin{array}{lll}
  V_\theta= U\sin \theta+(V_0+V)\cos \theta, \\
       V_R=-U\cos \theta+(V_0+V)\sin \theta,
 \label{VRVT}
 \end{array}
 \end{equation}
where the position angle $\theta$ satisfies the relation
$\tan \theta=y/(R_0-x);$ $x, y,$ and $z$ are the rectangular
heliocentric coordinates of the star (the velocities $U,
V,$ and $W$ are directed along the corresponding $x,
y,$ and $z$ axes); and $V_0$ is the linear Galactic rotation
velocity at the solar distance $R_0.$

Here, to take into account the influence of the
spiral density wave, we use the simplest model based
on the linear theory of density waves by Lin and
Shu (1964), in which the potential perturbation is in
the form of a traveling wave. Then,
  \begin{equation}
 \begin{array}{lll}
 \tilde{v}_R = f_R \cos \chi,\qquad
 \tilde{v}_\theta = f_\theta \sin \chi,\qquad
 \chi= m[{\rm \cot} (i)\ln (R/R_0)-\theta]+\chi_\odot,
 \label{wave}
 \end{array}
 \end{equation}
where $f_R$ and $f_\theta$ are the amplitudes of the radial
(directed toward the Galactic center in the arm) and
azimuthal (directed along the Galactic rotation) velocity
perturbations; $i$ is the spiral pitch angle ($i<0$
for winding spirals); $m$ is the number of arms; $\chi_\odot$ is
the phase angle of the Sun, in this paper we measure
it from the center of the Carina–Sagittarius arm; $\lambda,$
the distance (along the Galactocentric radial direction)
between adjacent segments of the spiral arms in
the solar neighborhood (the wavelength of the spiral
density wave), is calculated from the relation
 \begin{equation}
 {\rm \tan } (i)=\lambda m/(2\pi R_0).
 \label{tani}
 \end{equation}
The presented method of allowance for the influence
of the spiral density wave was used, for example,
by Mishurov and Zenina (1999) or Fern\'andez
et al. (2001), where its detailed description can be
found.

It can be seen that in a small solar neighborhood,
as in our case, the position angle $\theta\rightarrow0^\circ$ in Eq. (6)
and, therefore, allowance for the spiral density wave
does not depend on m. According to the analysis
of various stellar samples (Dambis et al. 2015;
Rastorguev et al. 2017; Bobylev and Bajkova 2019a;
Loktin and Popova 2019), in this paper we adopt
the following parameters of the spiral density wave:
$\lambda=2.2$ kpc, $f_R=5$ km s$^{-1}$, $f_\theta=0$ km s$^{-1}$, and
$\chi_\odot=-120^\circ.$

 \subsection*{Residual Velocity Ellipsoid}
To determine the parameters of the stellar residual velocity ellipsoid, we use the following well-known method (Trumpler and Weaver 1953; Ogorodnikov 1965). In the classical case, six second-order moments $a, b, c, f, e,$ and $d$ are considered:
\begin{equation}
 \begin{array}{lll}
 a=\langle U^2\rangle-\langle U^2_\odot\rangle,\qquad\quad
 b=\langle V^2\rangle-\langle V^2_\odot\rangle,\qquad\quad
 c=\langle W^2\rangle-\langle W^2_\odot\rangle,\\
 f=\langle VW\rangle-\langle V_\odot W_\odot\rangle,\quad
 e=\langle WU\rangle-\langle W_\odot U_\odot\rangle,\quad
 d=\langle UV\rangle-\langle U_\odot V_\odot\rangle,
 \label{moments}
 \end{array}
 \end{equation}
However, as has been noted above, the observed velocities
can be freed not only from the peculiar solar
motion, but also from other effects. The moments
$a, b, c, f, e,$ and $d$ are the coefficients of the surface
equation
 \begin{equation}
 ax^2+by^2+cz^2+2fyz+2ezx+2dxy=1,
 \end{equation}
and the components of the symmetric residual velocity
moment tensor
 \begin{equation}
 \left(\matrix {
  a& d & e\cr
  d& b & f\cr
  e& f & c\cr }\right).
 \label{ff-5}
 \end{equation}
All elements of this tensor can be determined by
solving the following system of conditional equations:
\begin{equation}
 \begin{array}{lll}
 V^2_l= a\sin^2 l+b\cos^2 l\sin^2 l-2d\sin l\cos l,
 \label{EQsigm-2}
 \end{array}
 \end{equation}
\begin{equation}
 \begin{array}{lll}
 V^2_b= a\sin^2 b\cos^2 l+b\sin^2 b\sin^2 l+c\cos^2 b\\
 -2f\cos b\sin b\sin l-2e\cos b\sin b\cos l+2d\sin l\cos l\sin^2 b,
 \label{EQsigm-3}
 \end{array}
 \end{equation}
\begin{equation}
 \begin{array}{lll}
 V_lV_b= a\sin l\cos l\sin b+b\sin l\cos l\sin b\\
 +f\cos l\cos b-e\sin l\cos b+d(\sin^2 l\sin b-\cos^2\sin b),
 \label{EQsigm-4}
 \end{array}
 \end{equation}
\begin{equation}
 \begin{array}{lll}
 V_b V_r=-a\cos^2 l\cos b\sin b-b\sin^2 l\sin b\cos b+c\sin b\cos b\\
 +f(\cos^2 b\sin l-\sin l\sin^2 b)+e(\cos^2 b\cos l-\cos l\sin^2 b)\\
 -d(\cos l\sin l\sin b\cos b+\sin l\cos l\cos b\sin b),
 \label{EQsigm-5}
 \end{array}
 \end{equation}
\begin{equation}
 \begin{array}{lll}
 V_l V_r=-a\cos b\cos l\sin l+b\cos b\cos l\sin l\\
    +f\sin b\cos l-e\sin b\sin l+d(\cos b\cos^2 l-\cos b\sin^2 l).
 \label{EQsigm-6}
 \end{array}
 \end{equation}
Its solution is sought by the least-squares method for
the six unknowns $a, b, c, f, e,$ and $d$. The eigenvalues
of the tensor (10) $\lambda_{1,2,3}$ are then found from the
solution of the secular equation
 \begin{equation}
 \left|\matrix
 {a-\lambda&        d&        e\cr
       d & b-\lambda &        f\cr
       e &          f&c-\lambda\cr }
 \right|=0.
 \label{ff-7}
 \end{equation}
The eigenvalues of this equation are equal to the reciprocals
of the squares of the semiaxes of the velocity
moment ellipsoid and, at the same time, the squares
of the semiaxes of the residual velocity ellipsoid:
 \begin{equation}
 \begin{array}{lll}
 \lambda_1=\sigma^2_1, \lambda_2=\sigma^2_2, \lambda_3=\sigma^2_3,\qquad
 \lambda_1>\lambda_2>\lambda_3.
 \end{array}
 \end{equation}
The directions of the principal axes of the tensor (16) $L_{1,2,3}$ and
$B_{1,2,3}$ are found from the relations
 \begin{equation}
 \tan L_{1,2,3}={{ef-(c-\lambda)d}\over {(b-\lambda)(c-\lambda)-f^2}},
 \label{ff-41}
 \end{equation}
 \begin{equation}
 \tan B_{1,2,3}={{(b-\lambda)e-df}\over{f^2-(b-\lambda)(c-\lambda)}}\cos L_{1,2,3}.
 \label{ff-42}
 \end{equation}
The errors in $L_{1,2,3}$ and $B_{1,2,3}$ are estimated as follows:
 \begin{equation}
 \renewcommand{\arraystretch}{2.2}
  \begin{array}{lll}
  \displaystyle
 \varepsilon (L_2)= \varepsilon (L_3)= {{\varepsilon (\overline {UV})}\over{a-b}},\\
  \displaystyle
 \varepsilon (B_2)= \varepsilon (\varphi)={{\varepsilon (\overline {UW})}\over{a-c}},\\
  \displaystyle
 \varepsilon (B_3)= \varepsilon (\psi)= {{\varepsilon (\overline {VW})}\over{b-c}},\\
  \displaystyle
 \varepsilon^2 (L_1)={\varphi^2 \varepsilon^2 (\psi)+\psi^2 \varepsilon^2 (\varphi)\over{(\varphi^2+\psi^2)^2}},\\
  \displaystyle
 \varepsilon^2 (B_1)= {\sin^2 L_1 \varepsilon^2 (\psi)+\cos^2 L_1 \varepsilon^2 (L_1)\over{(\sin^2 L_1+\psi^2)^2}},
 \label{ff-65}
  \end{array}
 \end{equation}
where $ \varphi={\rm \cot} B_1 \cos L_1$ and $\psi={\rm \cot} B_1 \sin L_1.$ In
this case, the three quantities $\overline {U^2V^2}$, $\overline {U^2W^2}$ and $\overline {V^2W^2},$ should be calculated in advance. Then,
 \begin{equation}
 \renewcommand{\arraystretch}{1.6}
  \begin{array}{lll}
  \displaystyle
 \varepsilon^2 (\overline {UV})= (\overline{U^2V^2}-d^2)/n, \\
  \displaystyle
 \varepsilon^2 (\overline {UW})= (\overline {U^2W^2}-e^2)/n, \\
  \displaystyle
 \varepsilon^2 (\overline {VW})= (\overline {V^2W^2}-f^2)/n,
 \label{ff-73}
  \end{array}
 \end{equation}
where $n$ is the number of stars. Here, the errors of each axis are estimated by an independent method, except for $L_2$ and $L_3,$ whose errors are calculated from
the same formula.

Based on this approach, Bobylev and Bajkova (2017) studied the kinematic properties of protoplanetary nebulae, while Bobylev and Bajkova (2019b) analyzed the properties of the residual velocity ellipsoid for hot subdwarfs from the Gaia DR2 catalogue, where only
three Eqs. (11)--(13) were used, because there was not information about the line-of-sight velocities of such stars.

 \subsection*{Position Ellipsoid}
Let $m,n,k$ be the direction cosines of the pole
of the sought-for great circle from the $x, y,$ and $z$ axes.
The sought-for symmetry plane of the stellar system
is then determined as the plane for which the sum of
the squares of the heights, $h=mx+ny+kz$, is at a
minimum:
 \begin{equation}
 \sum h^2=\hbox {min}.
 \label{fff-2}
 \end{equation}
The sum of the squares $h^2=x^2m^2+y^2n^2+z^2k^2+2yznk+2xzkm+2xymn$ can be designated as
$2P=\sum h^2.$ As a result, the problem is reduced to searching for the minimum of the function $P:$
 \begin{equation}
 2P=Am^2+Bn^2+Ck^2+2Fnk+2Ekm+2Dmn,
 \label{fff-4}
 \end{equation}
where the second-order moments of the coordinates
$A=[xx],$
 $B=[yy],$
 $C=[zz],$
 $F=[yz],$
 $E=[xz],$
 $D=[xy],$ written via the Gauss brackets, are the
components of a symmetric tensor:
 \begin{equation}
 \left(\matrix {
  A& D & E\cr
  D& B & F\cr
  E& F & C\cr }\right),
 \label{fff-5}
 \end{equation}
whose eigenvalues $\lambda_{1,2,3}$ are found from the solution
of the secular equation
 \begin{equation}
 \left|\matrix
 {A-\lambda&        D&        E\cr
       D & B-\lambda &        F\cr
       E &          F&C-\lambda\cr }
 \right|=0.
 \label{fff-7}
 \end{equation}
The directions of the principal axes, $l_{1,2,3}$ and $b_{1,2,3}$ are determined similarly to the approach (18), (19) described above:
 \begin{equation}
 {\rm \tan}~l_{1,2,3}={{EF-(C-\lambda)D}\over {(B-\lambda)(C-\lambda)-F^2}},
 \label{fff-41}
 \end{equation}
 \begin{equation}
 {\rm \tan}~b_{1,2,3}={{(B-\lambda)E-DF}\over{F^2-(B-\lambda)(C-\lambda)}}\cos l_{1,2,3}.
 \label{fff-42}
 \end{equation}
The relations to estimate the errors in $l_{1,2,3}$ and $b_{1,2,3}$ are analogous to (20) and (21), where instead of the velocities$\overline {UV},$ $\overline {UW},$ $\overline {VW},$ $\overline {U^2V^2}$, $\overline {U^2W^2}$ and $\overline {V^2W^2},$ the corresponding coordinates $\overline {xy},$ $\overline {xz},$ $\overline {yz},$ $\overline
{x^2y^2}$, $\overline {x^2z^2}$ and $\overline {y^2z^2}$ should be used.

Thus, the algorithm for solving the problem consists in (i) setting up the function $2P$ (23), (ii) seeking for the roots of the secular equation (25), and (iii) estimating
the directions of the principal axes of the position ellipsoid $l_{1,2,3}$ and $b_{1,2,3}$. Based on this approach, for example, using masers with measured trigonometric parallaxes, Bobylev and Bajkova (2014) redetermined the spatial orientation parameters of the Local arm.

 \begin{figure}[t] {\begin{center}
 \includegraphics[width=140mm]{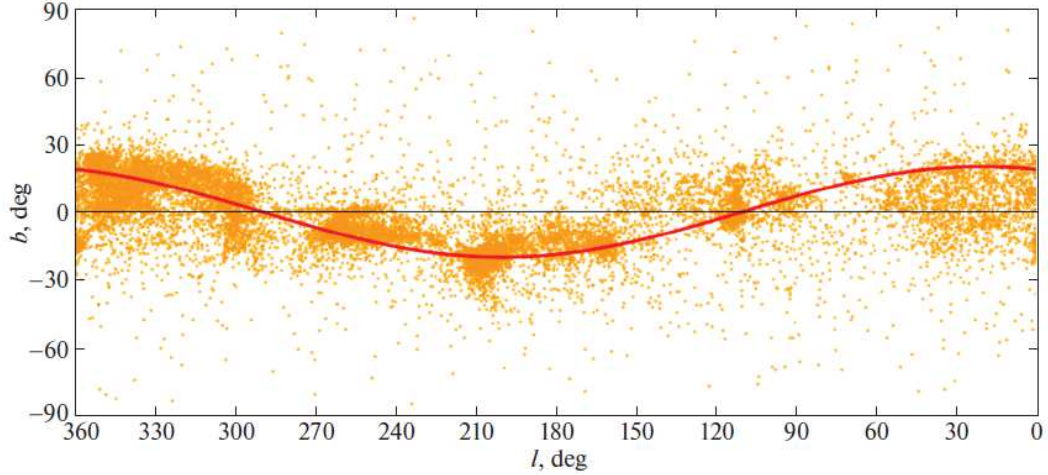}
 \caption{
(Color online) Distribution of the PMS3 stars on the celestial sphere; the solid line indicates a cosine wave with an
amplitude of 20$^\circ$. }
 \label{f-lb}
 \end{center} } \end{figure}

 \subsection*{Kinematic Model}
From an analysis of the residual velocities $V_r,V_l,V_b$ we can determine the mean group velocity $U_G,V_G,W_G,$  and four analogs of the Oort constants $A_G,B_G,C_G,K_G$ ($G$ is the Gould Belt), which, in our case, characterize the intrinsic rotation ($A_G$ and $B_G$) and expansion/contraction ($C_G$ and $K_G$) of the sample of low-mass stars closely associated with the Gould Belt based on the simple Oort–Lindblad kinematic model:
 \begin{equation}
 \begin{array}{lll}
 V_r=  U_G\cos b\cos l
      +V_G\cos b\sin l
      +W_G\sin b\\
      +rA_G\cos^2 b\sin 2l+rC_G\cos^2 b \cos 2l +rK_G\cos^2 b,
 \label{MOD-1}
 \end{array}
 \end{equation}
 \begin{equation}
 \begin{array}{lll}
 V_l= -U_G\sin l
      +V_G\cos l\\
 +rA_G\cos b\cos 2l-rC_G\cos b\sin 2l+rB_G\cos b,
 \label{MOD-2}
 \end{array}
 \end{equation}
 \begin{equation}
 \begin{array}{lll}
 V_b=-U_G\cos l\sin b
     -V_G\sin l \sin b
     +W_G\cos b\\
    -rA_G\sin b\cos b\sin 2l-rC_G\cos b\sin b\cos 2l-rK_G\cos b\sin b.
 \label{MOD-3}
 \end{array}
 \end{equation}
We find the unknowns $U_G,V_G,W_G,$ and $A_G,B_G,C_G,K_G$ by simultaneously solving the system of conditional equations (28)--(30) by the least-squares method (LSM).

 \section*{RESULTS}
Let us first consider the simplest method of estimating the geometric parameters of the Gould Belt from the distribution of stars on the celestial sphere. Figure 2 presents such a distribution for the PMS3 stars. The cosine wave with an amplitude of $20^\circ$ and
a phase shift of $20^\circ$ is drawn in such a way that the longitude of the ascending node here is $l_\Omega=290^\circ$. The curve can also be fitted to the data more accurately.
However, we should take into account the fact that this is only a projection of stars located at various heliocentric distances onto the sphere. Therefore, an analysis of the spatial coordinates of stars should yield more objective results.

The parameters of the position ellipsoids for three samples of stars are given in Table 1. The principal semiaxes of the position ellipsoid are determined to within a constant. As can be seen from the table, the ellipsoid becomes increasingly elongated along the $x$
axis from PMS1 to PMS3. If the size of the first semiaxis is taken to be 350 pc, then the ellipsoid, for example, for the PMS3 sample will have sizes $350\times270\times87$ pc very close to those of the Lindblad ring ($350\times250\times50$ pc).

 \begin{table}[t]
 \caption[]{\small
Parameters of the position ellipsoids for three samples of stars with relative trigonometric parallax errors less than 15\%
 }
  \begin{center}  \label{t:1}   \small
  \begin{tabular}{|c|r|r|r|r|r|}\hline
  Parameters   &         pms1 &         pms2 &         pms3 \\\hline
  $N_\star$    &        43706 &        33978 &        23683 \\
  $\lambda_1$  & $55.7\pm0.1$ & $49.2\pm0.1$ & $40.4\pm0.1$ \\
  $\lambda_2$  & $47.7\pm0.1$ & $40.2\pm0.1$ & $31.3\pm0.1$ \\
  $\lambda_3$  & $17.7\pm0.1$ & $14.0\pm0.3$ & $10.0\pm0.1$ \\
  $\lambda_1:\lambda_2:\lambda_3$ & $1:0.86:0.32$ & $1:0.82:0.29$ & $1:0.77:0.25$ \\

  $l_1, b_1$ & $~35.7\pm0.5^\circ,$ $~11.0\pm0.1^\circ$ & $~38.6\pm0.7^\circ,$ $12.1\pm0.2^\circ$ & $~35.0\pm1.2^\circ,$ $14.3\pm0.6^\circ$ \\
  $l_2, b_2$ & $125.4\pm0.5^\circ,$ $~-2.0\pm0.1^\circ$ & $127.9\pm0.9^\circ,$ $-3.1\pm0.1^\circ$ & $124.5\pm0.6^\circ,$ $-2.0\pm0.2^\circ$ \\
  $l_3, b_3$ & $206.1\pm0.5^\circ,$ $~78.9\pm0.1^\circ$ & $203.8\pm0.9^\circ,$ $77.5\pm0.1^\circ$ & $206.8\pm0.6^\circ,$ $75.6\pm0.3^\circ$ \\
  \hline
  \end{tabular}\end{center} \end{table}
 \begin{table}[t]
 \caption[]{\small
Parameters of the residual velocity ellipsoids for three samples of stars using only their propermotions (the upper part of the table) and in the simultaneous solution with the addition of line-of-sight velocities whose errors do not exceed 2 km s$^{-1}$ (the lower part of the table)
 }
  \begin{center}  \label{t:2}   \small
  \begin{tabular}{|c|r|r|r|r|r|}\hline
  Parameters    &         pms1 &         pms2 &         pms3 \\\hline
  $N_\star$    &        43706 &        33978 &        23683 \\
  $\sqrt{\sigma_0},$ km s$^{-1}$ & $8.7$ & $7.7$ & $6.9$ \\

  $\sigma_1,$ km s$^{-1}$ & $10.20\pm0.09$ & $9.55\pm0.10$ & $8.70\pm0.12$ \\
  $\sigma_2,$ km s$^{-1}$ & $ 7.21\pm0.12$ & $5.96\pm0.16$ & $4.72\pm0.23$ \\
  $\sigma_3,$ km s$^{-1}$ & $ 3.86\pm0.04$ & $3.36\pm0.05$ & $2.93\pm0.05$ \\

  $L_1, B_1$ & $~90\pm2^\circ,$ $~~2\pm0^\circ$ & $~98\pm4^\circ,$ $~~3\pm1^\circ$ & $111\pm7^\circ,$ $~~3\pm1^\circ$ \\
  $L_2, B_2$ & $179\pm1^\circ,$  $-8\pm1^\circ$ & $187\pm1^\circ,$ $-15\pm1^\circ$ & $200\pm2^\circ,$ $-31\pm2^\circ$ \\
  $L_3, B_3$ & $196\pm1^\circ,$ $~82\pm1^\circ$ & $199\pm1^\circ,$ $~74\pm1^\circ$ & $206\pm2^\circ,$ $~59\pm1^\circ$ \\
  \hline
  $N_\star$    &        41081 &        32125 &        22480 \\
  $\sqrt{\sigma_0},$ km s$^{-1}$ & $9.7$ & $8.5$ & $7.5$ \\

  $\sigma_1,$ km s$^{-1}$ & $10.58\pm0.08$ & $9.76\pm0.08$ & $8.87\pm0.10$ \\
  $\sigma_2,$ km s$^{-1}$ & $ 8.47\pm0.10$ & $7.04\pm0.14$ & $5.58\pm0.20$ \\
  $\sigma_3,$ km s$^{-1}$ & $ 3.79\pm0.03$ & $3.33\pm0.03$ & $3.03\pm0.04$ \\

  $L_1, B_1$ & $~77\pm1^\circ,$ $~~4\pm0^\circ$ & $~95\pm2^\circ,$ $~~~3\pm1^\circ$ & $112\pm3^\circ,$ $~~~2\pm0^\circ$ \\
  $L_2, B_2$ & $167\pm2^\circ,$  $-6\pm1^\circ$ & $184\pm2^\circ,$ $-12\pm1^\circ$ & $202\pm2^\circ,$ $-22\pm1^\circ$ \\
  $L_3, B_3$ & $197\pm2^\circ,$ $~83\pm1^\circ$ & $200\pm2^\circ,$ $~~78\pm1^\circ$ & $208\pm2^\circ,$ $~~68\pm1^\circ$ \\
  \hline
  \end{tabular}\end{center} \end{table}

 \begin{table}[t]
 \caption[]{\small
The parameters of the Oort–Lindblad kinematic model that we found based on two samples of stars only from the stars with line-of-sight velocities (upper part) and from all data (lower part)
 }
  \begin{center}  \label{t:3}   \small
  \begin{tabular}{|c|r|r|r|r|r|r|r|}\hline
  Parameters       &\multicolumn{2}{c|}{ums}  &\multicolumn{2}{c|}{pms3} \\\hline
  & before correction & after correction & before correction & after correction \\\hline
  $N_\star$        &        13092 &        13092 &         1877 &         1877 \\
  $\sigma_0,$ km s$^{-1}$ &        12.7  &        12.4  &        10.1  &         9.9  \\
   $U_\odot/U_G,$ km s$^{-1}$ & $ 6.43\pm0.11$ & $ 6.52\pm0.11$ & $ 5.25\pm0.24$ & $ 7.56\pm0.24$ \\
   $V_\odot/V_G,$ km s$^{-1}$ & $ 8.23\pm0.11$ & $ 3.95\pm0.11$ & $11.82\pm0.24$ & $-0.18\pm0.24$ \\
   $W_\odot/W_G,$ km s$^{-1}$ & $ 7.22\pm0.11$ & $-0.21\pm0.11$ & $ 5.36\pm0.23$ & $ 1.14\pm0.23$ \\

   $V,$ km s$^{-1}$ & $12.70\pm0.11$ & $7.63\pm0.11$ & $14.00\pm0.24$ & $ 7.65\pm0.24$ \\
   $l,$ deg.  & $ 52\pm1$      & $  31\pm1$    & $   66\pm1$    & $   9\pm2$    \\
   $b,$ deg.  & $ 35\pm1$      & $  -2\pm1$    & $   23\pm1$    & $   9\pm3$    \\

   $A,$ km s$^{-1}$ kpc$^{-1}$ & $12.01\pm0.35$ & $-3.50\pm0.35$ & $ 14.38\pm0.96$ & $ -0.64\pm0.95$ \\
   $B,$ km s$^{-1}$ kpc$^{-1}$ & $-7.89\pm0.35$ & $ 3.39\pm0.33$ & $-16.48\pm0.92$ & $ -4.21\pm0.91$ \\
   $C,$ km s$^{-1}$ kpc$^{-1}$ & $-2.78\pm0.35$ & $-7.18\pm0.35$ & $ -4.02\pm0.95$ & $ -9.17\pm0.94$ \\
   $K,$ km s$^{-1}$ kpc$^{-1}$ & $ 6.42\pm0.37$ & $ 0.77\pm0.36$ & $  5.76\pm1.00$ & $  0.12\pm0.99$ \\

 $l_{xy},$ deg. & $ 5\pm1$       &      $-32\pm1$ &  $ 8\pm2$       &     $-43\pm3$ \\
  \hline

  $N_\star$        &        71594 &        71594 &        23668 &        23668 \\
  $\sigma_0,$ km s$^{-1}$ &        10.1  &         9.7  &         4.3  &         3.9  \\

   $U_\odot/U_G,$ km s$^{-1}$ & $ 7.56\pm0.05$ & $ 5.35\pm0.05$ & $10.30\pm0.05$ & $ 2.55\pm0.04$ \\
   $V_\odot/V_G,$ km s$^{-1}$ & $ 8.36\pm0.05$ & $ 3.66\pm0.05$ & $12.61\pm0.04$ & $-0.62\pm0.04$ \\
   $W_\odot/W_G,$ km s$^{-1}$ & $ 7.03\pm0.04$ & $-0.09\pm0.04$ & $ 6.02\pm0.03$ & $ 0.79\pm0.03$ \\

   $V,$ km s$^{-1}$ & $13.28\pm0.05$ & $ 6.49\pm0.05$ & $17.35\pm0.04$ & $ 2.75\pm0.04$ \\
   $l,$ deg.  & $ 47.9\pm0.2$ & $ 34.4\pm0.4$ & $  5.8\pm0.2$  & $  346\pm2$    \\
   $b,$ deg.  & $ 32.0\pm0.2$ & $ -0.8\pm0.3$ & $ 20.3\pm0.1$  & $   17\pm1$    \\

   $A,$ km s$^{-1}$ kpc$^{-1}$ & $ 10.45\pm0.15$ & $-4.84\pm0.14$ & $  7.01\pm0.12$ & $ -7.20\pm0.11$ \\
   $B,$ km s$^{-1}$ kpc$^{-1}$ & $-11.26\pm0.11$ & $ 0.32\pm0.11$ & $-16.72\pm0.10$ & $ -3.83\pm0.09$ \\
   $C,$ km s$^{-1}$ kpc$^{-1}$ & $ -3.70\pm0.15$ & $-8.68\pm0.14$ & $  4.61\pm0.15$ & $  0.22\pm0.14$ \\
   $K,$ km s$^{-1}$ kpc$^{-1}$ & $  2.98\pm0.26$ & $-2.48\pm0.25$ & $  7.03\pm0.32$ & $  1.31\pm0.29$ \\

 $l_{xy},$ deg. & $ 10\pm1$       & $-30\pm1$      &  $-17\pm1$      &     $1\pm1$ \\
  \hline
  \end{tabular}\end{center} \end{table}

We can judge the inclination typical for the Gould Belt by the angles $b_1$ and $b_3.$ It can be seen from Table 1 that the inclinations deduced from the PMS1 and PMS2 samples are small, 11--12$^\circ$, they are quite far from the expected values. This suggests that the
samples are contaminated by background stars and that it is difficult to separate two layers of stars---the layer of Gould Belt nonmembers lying in the Galactic plane and the inclined layer of Gould Belt members. The inclination of 14$^\circ$ found from the PMS3 sample
is not very large either. The position of the third axis of the ellipsoid allows the longitude of the ascending node of the PMS3 stellar system to be determined,
$l_\Omega=l_3+90^\circ=297\pm1^\circ$.

Table 2 gives the parameters of the residual velocity
ellipsoids for three samples of stars. When forming
them, we took into account the peculiar solar motion
and the differential Galactic rotation in Eqs. (2)--(4).
The solution was obtained by two methods. The results
obtained only from the stellar proper motions are
given in the upper part of the table, while the results
obtained from the same stars, but the equation for
the line-of-sight velocity (whose errors do not exceed
2 km s$^{-1}$), if available, is also used, are given in the
lower part of the table.

When using only the stellar proper motions, we obtain the solution with the smallest errors of the parameters being determined. In this case, however, we slightly underestimate them, because we assumed the line-of-sight velocities to be zero when calculating
the velocities $U, V,$ and $W$ and the corresponding
errors (see Eqs. (20) and (21)). Therefore, the
results obtained by invoking the stellar line-of-sight
velocities should be deemed more reliable. The results
obtained from the PMS3 sample are of greatest
interest. In particular, note the position of the first
axis of the velocity ellipsoid $L_1=112\pm3^\circ,$ which is
closely related to the direction toward the kinematic
center. For example, if we are dealing with the intrinsic
rotation of the stellar system (in the absence of
intrinsic expansion), then $L_1$ should point exactly to
the expansion center. Conversely, in the presence of intrinsic expansion (and zero rotation) the direction $L_1$ will differ by 45$^\circ$ from the direction toward the
system's kinematic center (Ogorodnikov 1965).

Figure 3 presents the distribution of the PMS3 stars in the $xy, xz,$ and $yz$ planes. The stellar position ellipsoid is shown. Note that the mean coordinates
 $({\overline x},{\overline y},{\overline z})=(-65,-79,-35)$ pc calculated from the
entire PMS3 sample provide information about the concentration center of the sample stars. As can be seen from the figure, the ellipse center was moved
along the y axis and placed in the region of the lowest
concentration of stars. In this case, the direction
to the ellipse center is in good agreement with the
direction $L_1=112^\circ$ found by analyzing the residual
velocity ellipsoid as a presumed direction to the kinematic
center of the stellar system (Table 2). We clearly
see from Fig. 3b that it is desirable to impart a slightly
larger inclination to the ellipse. Thus, the inclination
$B_2=22\pm1^\circ$ found from our analysis of the residual
velocity ellipsoid is closer to the value typical for the
Gould Belt

To estimate the effects of intrinsic rotation and expansion/contraction of the stellar systems under consideration, we solve the system of conditional equations (28)--(30) by the LSM. We seek its solution based on two samples, UMS and PMS3, both without
and with allowance for the influence of the spiral density wave.

The results are presented in Table 3. The model parameters are given in the first column, in the second and fourth columns the velocities were not freed from any effects, and in the third and fifth columns the stellar velocities were freed from the solar motion, the
differential Galactic rotation, and the influence of the spiral density wave. Using the parameters $A$ and $C$ found, we calculated the angle $l_{xy}$ (vertex deviation)
according to the well-known relation
 $${\rm \tan} (2l_{xy})=-C/A,$$
which is valid in the absence of expansion. This angle specifies the direction to the kinematic center of the stellar system.

 \begin{figure} [t] {\begin{center}
 \includegraphics[width=150mm]{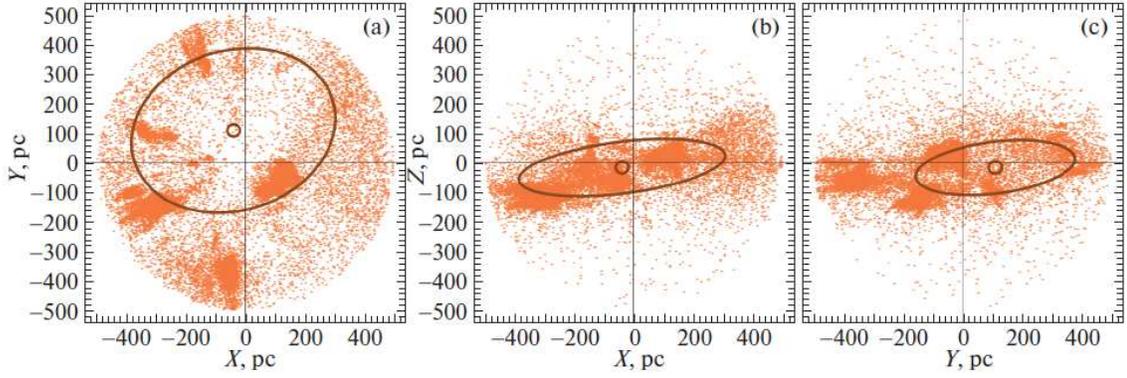}
 \caption{
(Color online) Spatial distribution of the PMS3 stars; the ellipsoid was found by analyzing the positions of these stars. }
 \label{f-XY-33}
 \end{center} } \end{figure}

The kinematic equations (28)--(30) were solved by two methods. The results obtained from the stars with complete information, i.e., the parallax, two proper motion components, and line-of-sight velocity are known for each star, are presented in the upper part of
Table 3, while the results obtained from the stars with incomplete information, i.e., only the stellar proper motions were used in the absence of line-of-sight velocities, are presented in the lower part of Table 3.

Since the velocities in the second and fourth columns are free from the corrections, the velocities $U_\odot, V_\odot$ and $W_\odot$ have the ordinary meaning of the sample group velocity. In contrast, the solar velocity relative to the local standard of rest (LSR) with the values from Sch\"onrich et al. (2010), ($(U,V,W)_\odot=(11.1,12.2,7.3)$ km s$^{-1}$, was taken into account in the third and fifth columns; therefore, $U_G, V_G, W_G,$ $V,$ $l$ and $b$ show the motion of the entire sample relative to the LSR. Here, the velocity is
$V=\sqrt{U^2_G+V^2_G+W^2_G}$ and its direction is $l_G$ and $b_G.$ The values of these quantities strongly depend on the adopted peculiar solar velocity relative to the LSR. For example, based on open star clusters younger than 60 Myr from the Gould Belt, Bobylev (2004) found $(U,V,W)_G=(1.1,-11.8,1.3)$~km s$^{-1}$, $l_G=275^\circ$ and $b=6^\circ,$, where the peculiar solar velocity components $(U,V,W)_\odot=(10.0,5.3,7.2)$ km s$^{-1}$ from Dehnen and Binney (1998) were used.

Similarly, in the second and fourth columns the Oort constants $A$ and $B,$ to a lesser degree $C$ and $K,$ describe the differential Galactic rotation, while in the third and fifth columns these parameters already reflect exclusively the intrinsic kinematic properties of the sample stars.

An analysis of the results in Table 3 shows that allowance for the spiral density wave removes almost completely the positive $K$ effect (the expansion of the stellar system). Furthermore, a positive intrinsic rotation of the system with the angular velocity $B-A$
is observed in the residual stellar velocities (the third and fifth columns) and only from the PMS3 sample; in the upper part of the table this rotation is negative (i.e., it coincides in direction with the Galactic one) and has $B-A=-3.57\pm1.34$ km s$^{-1}$ kpc$^{-1}$. Strictly speaking (Ogorodnikov 1965), it should be slightly different, because there is a large value of the constant C here; therefore, $A_{\rm new}=\sqrt{A^2+C^2},$ and
then $B-A_{\rm new}=5.0\pm1.6$ km s$^{-1}$ kpc$^{-1}$. Thus, the rotation will be positive for the direction to the center $-43^\circ$ if in Eqs. (28)--(30) we substitute $l-(-43^\circ)$ for $l.$ The direction $l_{xy}=-43\pm3^\circ$ here can be interpreted as the fact that the direction to the center of the stellar system is on the line with longitudes $317-137^\circ$, with the direction $l=137^\circ$ pointing to the second Galactic quadrant, where the center of the Gould Belt is most likely located. In contrast, in the lower part of the table the intrinsic rotation of the PMS3 sample is positive with $B-A=3.37\pm0.14$ km s$^{-1}$ kpc$^{-1}$ at an almost zero value of the constant $C.$

The results from Table 3 obtained from the young massive stars of the UMS sample are of interest. There is a positive $K$ effect in their uncorrected velocities,
which is removed after allowance for the spiral
density wave. For the stars of this sample the errors
per unit weight $\sigma_0$ are noticeably larger. Since in the
catalogue by Zari et al. (2018) the stars were selected
with a fairly strong restriction on the absolute value of
the tangential velocity, $V_t<40$ km s$^{-1}$, the Galactic
rotation parameters ($A$ and $B$ in the second column
of the table) may be underestimated.

 \begin{figure} [t] {\begin{center}
 \includegraphics[width=150mm]{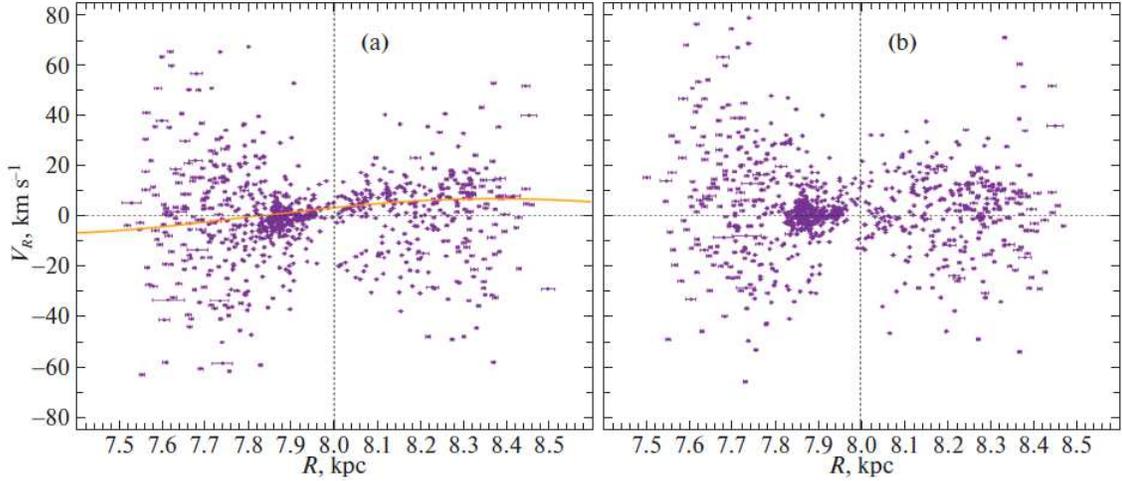}
 \caption{
(Color online) Galactocentric radial velocities, $V_R,$ of the PMS3 stars versus distance $R$ corrected for the solar motion (a) and for the Galactic rotation and the spiral density wave (b). }
 \label{f-Keffekt}
 \end{center} } \end{figure}

As we see from Fig. 1, the stars with line-of-sight
velocities in both samples under consideration occupy
slightly different regions on the H--R diagram compared
to the entire sample. Thus, they have a slightly
different evolutionary status. Most likely, there is a
significant fraction of type-A stars among the stars with measured line-of-sight velocities in the UMS sample (the dark circles in Fig. 1a). In contrast, there
is a large fraction of faint stars in the PMS3 sample
without line-of-sight velocities (the light circles in
Fig. 1b), where the parallax and proper motion errors
increase significantly compared to the brighter stars
from the Gaia DR2 catalogue. From this viewpoint,
the differences in kinematic parameters between the
upper and lower parts of the table should come as no
surprise.

As a result, in our opinion, the most reliable values of the derived kinematic parameters are contained in the fourth and fifth columns of the upper part of Table. 3. These parameters were deduced from the PMS3 sample. It is also interesting to note that the group velocity of this sample shows a close association with the Gould Belt. Indeed, as can be seen from the last column of the table, the velocity $V$ has a direction $l$ from 179$^\circ$ to 166$^\circ$ and $b$ from $-9^\circ$ to $-17^\circ$, i.e., it lies virtually in the plane of the Gould Belt.

Figure 4 presents the Galactocentric radial velocities $V_R$ of the PMS3 stars. In the first case (Fig. 4a), they were corrected only for the solar motion. An inclined arrangement of points is clearly seen in the immediate solar neighborhood with a radius of about
200 pc. This graph shows the wave
 $$
  -7\cos\biggl[-{2\pi R_0\over 2.2}\ln\biggl({R\over R_0}\biggr)-120^\circ\biggr],
 $$
written according to relations (6) and (7), with a perturbation amplitude $f_R=7$ km s$^{-1}$, a wavelength $\lambda=2.2$ kpc, and the Sun’s phase in the wave $\chi_\odot=-120^\circ.$ Here the minus in front of the formula means that at the center of the spiral arm (for example, at $R\approx7.2$ kpc) the perturbation is directed toward the
Galactic center.

In the second case (Fig. 4b), the velocities corrected for the solar motion, the differential Galactic rotation, and the influence of the spiral density wave
are given. As can be seen from the figure, allowance for all these effects makes the distribution of points horizontal. Interestingly, the densest clump of points
in Fig. 4 at $R\approx7.9$ kpc formed by stars from the Scorpio–Centaurus OB association also becomes more horizontal after allowance for the influence of the spiral density wave. However, the local inclination still remains, suggesting the presence of intrinsic
expansion of this association.

 \section*{DISCUSSION}
Based on OB stars from the Hipparcos catalogue (1997) younger than 60~Myr, Torra et al. (2000) determined the inclination, 16--22$^\circ$, and the longitude of the ascending node of the great circle, 275--295$^\circ$. Bobylev (2016) showed that the system of nearby
high-latitude molecular clouds could be fitted by an ellipse with sizes $350\times235\times140$ pc oriented at an angle of $17\pm2^\circ$ to the Galactic plane with a longitude of the ascending node of $337\pm1^\circ$. Since high-latitude
molecular clouds very far from the symmetry plane of the Gould Belt were considered, the third axis of this ellipsoid turned out to be unusually large.

Dzib et al. (2018) analyzed twelve star-forming regions containing young stars and closely associated with the Gould Belt. Kinematic data from the Gaia DR2 catalogue were used. They showed that this system could be fitted by an ellipsoid with sizes $358\times316\times70$ pc and the center in the second Galactic quadrant $(x,y,z)_0=(-82,39,-25)\pm(15,7,4)$ pc. The ellipsoid is oriented at an angle of $21\pm1^\circ$ to the Galactic plane with a longitude of the ascending node of $319\pm2^\circ$. A new estimate of the Gould Belt expansion velocity was also obtained from these data, $2.5\pm0.1$ km s$^{-1}$.

Having analyzed a large sample of clump giants from the Gaia DR2 catalogue, Gontcharov (2019) determined the inclination of the dust layer associated with the Gould Belt, $18\pm2^\circ$. In addition, he estimate the scale height of this dust layer to be
$170\pm40$ pc. Thus, the geometric characteristics of the PMS3 stars found in this paper (an inclination of $14-22^\circ$ and a longitude of the ascending node of the
great circle $297-298^\circ$) are in good agreement with the characteristics of the Gould Belt determined by various authors from other data. This suggests that
the overwhelming majority of young T Tauri stars from the PMS3 sample belong to the Gould Belt structure.

It is interesting to estimate the $K$ effect in angular units. By definition, $2K=V_R/R+\partial V_R/\partial R$ if the rotation velocity $V_\theta$ is independent of the angle $V_\theta,$ $\partial V_\theta/\partial\theta=0$ (Ogorodnikov 1965). Then, at
a constant angular velocity (i.e., at $\partial V_R/\partial
R=0$) $\partial V_R/\partial R=0$ and $2K=V_R/R$.

From our examination of a wave similar to that in Fig. 4 we find $2K=2f_R/(\lambda/2)$ and, consequently, $K=4.5$ km s$^{-1}$ kpc$^{-1}$. In this case, we should take
into account the fact that the correction strongly depends on the Sun's phase $\chi_\odot$. Thus, even if the influence of the Galactic spiral density wave is
taken into account, the Gould Belt can have a slight
residual expansion. For example, having analyzed
OB stars from the Hipparcos catalogue younger than
30 Myr, Lindblad et al. (1997) obtained an estimate of $K=12$ km s$^{-1}$ kpc$^{-1}$. Based on OB stars younger than 60 Myr, Torra et al. (2000) found $K=7.1\pm1.4$ km s$^{-1}$ kpc$^{-1}$. Based on a sample of young stars, Bobylev (2004) found $K=8\pm2$ km s$^{-1}$ kpc$^{-1}$.

The probability that there is an intrinsic expansion
of the Scorpio–Centaurus association even
after allowance for the influence of the spiral density wave is great. For example, Blaauw (1964) found the expansion coefficient for it to be $K=50$ km s$^{-1}$ kpc$^{-1}$. Based on a sample of young stars with data from the Hipparcos catalogue, Bobylev
and Bajkova (2007) refined this coefficient, $K=46\pm8$ km s$^{-1}$ kpc$^{-1}$. As can be seen from Fig. 4, the stars of this association exert a strong influence
on the K estimate for the Gould Belt. When analyzing young massive multiple systems, Bobylev
and Bajkova (2013) noted a significant radial velocity
gradient $V_R/R\sim40$ km s$^{-1}$ kpc$^{-1}$ in the region of
the Scorpio–Centaurus association. They suggested that the influence of the spiral density wave should be eliminated before determining the intrinsic expansion
parameters of the Scorpio–Centaurus association.
Indeed, as can be clearly seen from Fig. 4, the spiral density wave and the velocities of the Scorpio–Centaurus association are almost parallel to one another; therefore, it is difficult to separate one effect from the other.

The residual velocity dispersions of the PMS3 stars are low, for example, $\sigma_0=7.5$ km s$^{-1}$ (Table 2),and the principal semiaxes of the residual velocity ellipsoid
$\sigma_{1,2,3}=(8.87,5.58,3.03)\pm(0.10,0.20,0.04)$ km s$^{-1}$
are comparable to the velocity dispersion of the gas clouds belonging to the Gould Belt, 1--5 km s$^{-1}$ (Galli et al. 2019).

 \section*{CONCLUSIONS}
We studied the spatial and kinematic properties of a large sample of young pre-main stars. For this purpose we used the catalogue by Zari et al. (2018)
containing more than 40 000 T Tauri stars with their
proper motions and parallaxes from the Gaia DR2 catalogue. We also considered the kinematic properties of a large (more than 80 000) sample of young stars (these are stars of spectral types O, B, and A) from this catalogue that occupy the upper part on the H--R diagram. The line-of-sight velocities are known for some of these stars.

We validated the hypothesis of Zari et al. (2018) that the stars belonging to the PMS3 sample have a very close spatial and kinematic association with
the Gould Belt. The following characteristics of the
position ellipsoid were estimated from the coordinates
of thePMS3 stars: its sizes are $350\times270\times87$ pc and
it is oriented at an angle of $14\pm1^\circ$ to the Galactic
plane with a longitude of the ascending node of $297\pm1^\circ$.

Our analysis of the motions of PMS3 stars showed that the residual velocity ellipsoid with principal semiaxes $\sigma_{1,2,3}=(8.87,5.58,3.03)\pm(0.10,0.20,0.04)$ km s$^{-1}$
is oriented at an angle of $22\pm1^\circ$ to the Galactic plane
with a longitude of the ascending node of $298\pm2^\circ$.

We showed that much ($\sim$5--7 km s$^{-1}$ kpc$^{-1}$) of
the expansion effect (K effect) typical for Gould Belt
stars could be explained by the influence of a Galactic
spiral density wave. When making allowance for the
influence of the spiral density wave, we took into account
only the radial velocity perturbation component
$f_R=5$ km s$^{-1}$ by assuming that it made a major
contribution when allowing for the $K$ effect. After
allowance for the peculiar solar motion relative to the
LSR, the differential Galactic rotation, and the density
wave in the residual stellar velocities, the intrinsic
expansion becomes very small or even is replaced by
contraction. At the same time, an intrinsic rotation
with an magnitude of 3--6 km s$^{-1}$ kpc$^{-1}$ manifests
itself, with the sign of this rotation being most likely
positive. This effect should be studied further in more
detail.

 \subsection*{ACKNOWLEDGMENTS}
I am grateful to the referee for the useful remarks
that contributed to an improvement of the paper.

 \subsection*{FUNDING}
This work was supported in part by Program KP19--270 of the Presidium of the Russian Academy
of Sciences ``Questions of the Origin and Evolution of the Universe with the Application of Methods of Ground-Based Observations and Space Research''.

 \bigskip \bigskip\medskip{\bf REFERENCES}{\small

 1. A. Blaauw, Ann. Rev. Astron. Astrophys. 2, 213 (1964).

 2. V. V. Bobylev, Astron. Lett. 30, 784 (2004).

 3. V. V. Bobylev, Astron. Lett. 32, 816 (2006).

 4. V. V. Bobylev, Astrophysics 57, 583 (2014).

 5. V. V. Bobylev and A. T. Bajkova, Astron. Lett. 33, 571 (2007).

 6. V. V. Bobylev and A. T. Bajkova, Astron. Lett. 39, 532
(2013).

7. V. V. Bobylev and A. T. Bajkova, Astron. Lett. 40, 783
(2014).

8. V. V. Bobylev, Astron. Lett. 42, 544 (2016).

9. V. V. Bobylev and A. T. Bajkova Astron. Lett. 43, 452
(2017).

10. V. V. Bobylev and A. T. Bajkova, Astron. Lett. 45, 208
(2019a).

11. V. V. Bobylev and A. T. Bajkova, Astron. Rep. 63, 932
(2019b).

12. A. G. A. Brown, A. Vallenari, T. Prusti, de Bruijne,
C. Babusiaux, C. A. L. Bailer-Jones, M. Biermann,
D.W. Evans, et al. (Gaia Collab.), Astron. Astrophys.
616, 1 (2018).

13. A. K. Dambis, L. N. Berdnikov, Yu. N. Efremov,
A. Yu. Knyazev, A. S. Rastorguev, E. V. Glushkova,
V. V. Kravtsov, D. G. Turner, D. J. Majaess, and R.
Sefako, Astron. Lett. 41, 489 (2015).

14. T. M. Dame, D. Hartmann, and P. Thaddeus, Astrophys.
J. 547, 792 (2001).

15. W. Dehnen and J. J. Binney, Mon. Not. R. Astron.
Soc. 298, 387 (1998).

 16. S. A. Dzib, L. Loinard, G. N. Ortiz-Le\'on, L. F. Rodriguez, and P. A. B. Galli, Astrophys. J. 867, 151 (2018).

17. Yu.N. Efremov, Sites of Star Formation inGalaxies
(Nauka, Moscow, 1989) [in Russian].

18. D. Fern\'andez, F. Figueras, and J. Torra, Astron. Astrophys.
372, 833 (2001).

19. P. A. B. Galli, L. Loinard, H. Bouy, L. M. Sarro,
G. N. Ortiz-Le\'on, S. A. Dzib, J. Olivares, M. Heyer,
et al., Astron. Astrophys. 630, 137 (2019).

 20. G. A. Gontcharov, Astron. Lett. 45, 605 (2019).

 21. C. C. Lin and F. H. Shu, Astrophys. J. 140, 646 (1964).

 22. P. O. Lindblad, Bull. Astron. Inst. Netherland 19, 34 (1967).

 23. P. O. Lindblad, J. Palou\v s, K. Loden, and L. Lindegren, in HIPPARCOS Venice’97, Ed. by B. Battrick (ESA Publ. Div., Noordwijk, 1997), p. 507.

 24. P. O. Lindblad, Astron. Astrophys. 363, 154 (2000).

 25. L. Lindegren, J. Hernandez, A. Bombrun, S. Klioner, U. Bastian, M. Ramos-Lerate, A. de Torres, H. Steidelmuller, et al. (Gaia Collab.), Astron. Astrophys. 616, 2 (2018).

 26. A. V. Loktin and M. E. Popova, Astrophys. Bull. 74, 270 (2019).

 27. Yu. N.Mishurov and I. A. Zenina, Astron. Astrophys. 341, 81 (1999).

 28. K. F. Ogorodnikov, Dynamics of Stellar Systems (Fizmatgiz, Moscow, 1965; Pergamon, Oxford, 1965).

29. C. A. Olano, Astron. Astrophys. 121, 295 (2001).

 30. C. A. Perrot and I. A. Grenier, Astron. Astrophys. 404, 519 (2003).

 31. A. E. Piskunov, N. V. Kharchenko, S. R\"oser, E. Schilbach, and R.-D. Scholz, Astron. Astrophys. 445, 545 (2006).

32. W.G. L. P\"oppel, Fundam. Cosmic Phys. 18, 1 (1997).

33. W. G. L. P\"oppel, ASP Conf. Ser. 243, 667 (2001).

 34. A. S. Rastorguev, M. V. Zabolotskikh, A. K. Dambis, N. D. Utkin, V. V. Bobylev, and A. T. Bajkova, Astrophys. Bull. 72, 122 (2017).

 35. M. J. Sartori, J. R. D. Lepine, andW. S. Dias, Astron. Astrophys. 404, 913 (2003).

 36. R. Sch\"onrich, J. Binney, and W. Dehnen, Mon. Not. R. Astron. Soc. 403, 1829 (2010).
 
 37. The HIPPARCOS and Tycho Catalogues, ESA SP--1200 (1997).

38. J. Torra, D. Fern\'andez, and F. Figueras, Astron. Astrophys.
359, 82 (2000).

39. C. A. O. Torres, R. Quast, C. H. F. Melo, and
M. F. Sterzik, Handbook of Star Forming Regions,
Vol. 2,, Vol. 5 of The Southern Sky ASPMonograph
Publications, Ed. by Bo Reipurth (ASP, San Francisco,
2008).

40. R. J. Trumpler and H. F. Weaver, Statistical Astronomy
(Univ. of Calif. Press, Berkely, 1953).

41. O. O. Vasilkova, Astron. Lett. 40, 59 (2014).

42. E.Zari, H. Hashemi, A.G. A. Brown, K. Jardine, and
P. T. de Zeeuw, Astron. Astrophys. 620, 172 (2018).

43. P. T. de Zeeuw, R. Hoogerwerf, J. H. J. de Bruijne,
A. G. A. Brown, and A. Blaauw, Astron. J. 117, 354
(1999).

  }
  \end{document}